\documentclass[useAMS,usenatbib,usegraphicx]{mn2e}

\usepackage{times,epsfig}

\def\gsim{\;\rlap{\lower 2.5pt
 \hbox{$\sim$}}\raise 1.5pt\hbox{$>$}\;}
\def\lsim{\;\rlap{\lower 2.5pt
   \hbox{$\sim$}}\raise 1.5pt\hbox{$<$}\;}
\def\micron{~$\mu\textrm{m}$ }
\def\micronend{$\mu\textrm{m}$}

\title[The Evolution of Warm ULIRGs]{The merger--driven evolution of warm infrared luminous galaxies}
\author[Younger et al.]{
Joshua\, D.\, Younger,$^{\! 1}$\thanks{Email: jyounger@cfa.harvard.edu}
Christopher\, C.\, Hayward,$^{\! 1}$
Desika\, Narayanan,$^{\! 1, 2}$\and
T.\, J.\, Cox,$^{\! 1, 3}$
Lars\, Hernquist,$^{\! 1}$
and Patrik Jonsson$^{\! 4}$
\vspace*{1mm}\\
$^1$ Harvard--Smithsonian Center for Astrophysics, 60 Garden Street, Cambridge, MA 02138 \\
$^2$ CfA Fellow \\
$^3$ Keck Fellow \\
$^4$ Department of Physics, University of California, Santa Cruz, CA 95064
}

\date{\fbox{\sc Draft dated: \today\ }}
\pagerange{000--000}

\begin{document}

\maketitle

\begin{abstract}

We present a merger--driven evolutionary model for the production of luminous (LIRGs) and ultraluminous infrared galaxies (ULIRGs) with warm IR colours.  Our results show that simulations of gas--rich major mergers including star formation, black hole growth, and feedback can produce warm (U)LIRGs.  We also find that while the warm evolutionary phase is associated with increased AGN activity, star formation alone may be sufficient to produce warm IR colours.  However, the transition can be suppressed entirely -- even when there is a significant AGN contribution -- when we assume a single--phase ISM, which maximizes the attenuation.  Finally, our evolutionary models are consistent with the  25--to--60\micron flux density ratio vs. $L_{HX}/L_{IR}$ relation for local LIRGs and ULIRGs, and predict the observed scatter in IR color at fixed $L_{HX}/L_{IR}$.  Therefore, our models suggest a cautionary note in the interpretation of warm IR colours: while associated with periods of active black hole growth, they are probably produced by a complex mix of star formation and AGN activity intermediate between the cold star formation dominated phase and the birth of a bright, unobscured quasar.

\end{abstract}

\begin{keywords}
galaxies: formation -- galaxies: evolution -- galaxies: starburst  -- methods: numerical -- infrared: galaxies
\end{keywords}

\section{Introduction}
\label{sec:intro}

With bolometric energy outputs rivaling bright quasars, luminous (LIRGs) and ultraluminous (ULIRGs) infrared (IR) galaxies are some of the most extreme objects in the Universe \cite[for a review, see][]{sanders1996}.  Although they make up a small fraction of the local IR luminosity density, at $z\gsim 1$ they become cosmologically important \citep{lefloch2005,magnelli2009}, contributing a significant fraction of the diffuse extragalactic IR background \citep[e.g.,][]{fixsen1998,pei1999} and up to half of the star formation in the early Universe \citep[e.g.,][]{blain1999}.  Over the past two decades, a number of authors have argued that these extreme systems represent an intermediate stage in the merger--driven transformation of gas--rich quiescent disks into luminous quasars \citep[e.g.][]{sanders1988a,hopkins2006}.   As a result, a detailed understanding of (U)LIRGs -- including the engine that powers their luminosity -- is essential to any successful theory of galaxy formation.

Based on a number of observational arguments, energy production in most (U)LIRGs appears to be dominated by star formation (SF) both locally \citep[e.g.,][]{genzel1998,lutz1998} and at high redshift \citep[e.g.,][]{alexander2005,pope2008b,younger2007,younger2008highres,younger2008.egsulirgs}.  However, the ``warm" IR colours seen in some systems \citep[$S_{\rm 25\mu m}/S_{\rm 60 \mu m}\gsim 0.27$:][]{deGrijp1985,deGrijp1987} are thought to betray a heavily obscured active galactic nucleus (AGN) that contributes significantly to the IR emission \citep[see also][]{risaliti2000}.  Therefore, warm ULIRGs may represent the final IR luminous stage immediately prior to the birth of a bright quasar \citep{sanders1988b}.

Recent theoretical work has shown that mergers can drive starbursts consistent with those seen in (U)LIRGs \citep{mihos1994,mihos1996,hopkins2006.sf}.  Furthermore, the radiative transfer calculations of \citet{chakrabarti2007} demonstrate that warm IR colours are associated with increased AGN activity.  However, since the starburst and AGN activity are both produced via the same merger--driven inflow, it is possible that both heat the dust distribution.  Here, we investigate the relative contribution of star formation and AGN activity to the production of warm (U)LIRGs.  This work is organized as follows: in \S~\ref{sec:methods} we summarize our methodology, in \S~\ref{sec:results} we present our results and discuss some potential interpretations, and in \S~\ref{sec:conclude} we conclude.

\section{Methodology}
\label{sec:methods}

The hydrodynamical simulations presented in this work were performed using {\sc Gadget--2} \citep{springel2005}, a smoothed particle hydrodynamics (SPH) code using the entropy conserving formalism of \citet{SpringelHernquist2002}.  We include the effects of gas dissipation via radiative cooling \citep[see][]{katz1996} and a sub--resolution model \citep{SpringelHernquist2003} for the multiphase interstellar medium (ISM) with a star formation (SF) timescale tuned to match the local Kennicutt--Schmidt relation \citep{schmidt1959,kennicutt1998}.  Finally, we include sink particles representing supermassive black holes (SMBH) that undergo Eddington--limited Bondi--Hoyle--Littleton accretion \citep[for a review, see][]{edgar2004} and release thermal feedback into the interstellar medium \citep{SDH2005}.  A number of previous studies have found that this model of AGN feedback, assuming a radiative efficiency of $\approx 10\%$ and thermal coupling efficiency of $\approx 5\%$, yields AGN light curves that simultaneously reproduce both the observed quasar luminosity function \citep{hopkins2006a,hopkins2007a} and the full range of observed scalings between the final SMBH mass and galaxy properties \citep{dimatteo2005,robertson2006a,hopkins2007theory,younger2008.smbh}.

The progenitor galaxies were constructed following the methods described in \citet{SDH2005}, and are similar to the Milky Way--like model in \citet{robertson2006b}: they are bulgeless disks embedded in a \citet{hernquist1990} dark matter halo of mass $M_{vir} = 1.4\times 10^{12}$ $M_\odot$ with concentration and spin consistent with cosmological N--body simulations \citep[$c_{vir}=9$ and $\lambda = 0.033$:][]{bullock2001,bett2007} and baryonic mass fraction $m_b=0.041$.  Each of the progenitor disk galaxies is given an initial gas fraction of $f_g=0.4$ by mass.  They are then placed on a zero--energy parabolic orbit, as motivated by cosmological N--body simulations \citep{benson2005,khochfar2006}.  In this work, we examine three different orbital configurations \citep[see also][]{cox2006}: orientation e (tilted prograde--prograde: $\theta_1=30$, $\phi_1=60$, $\theta_2=-30$, $\phi_2=45$), h (coplanar prograde--prograde: $\theta_1=0$, $\phi_1=0$, $\theta_2=0$, $\phi_2=0$), and g (tilted prograde--retrograde: $\theta_1=150$ $\phi_1=0$, $\theta_2=-30$, $\phi_1=45$).  These are the same merger simulations used by \citet{narayanan2008.winds} to study the effects of AGN-- and starburst-- driven winds on molecular gas emission from IR luminous galaxies. 

\begin{figure*}
\epsfig{figure=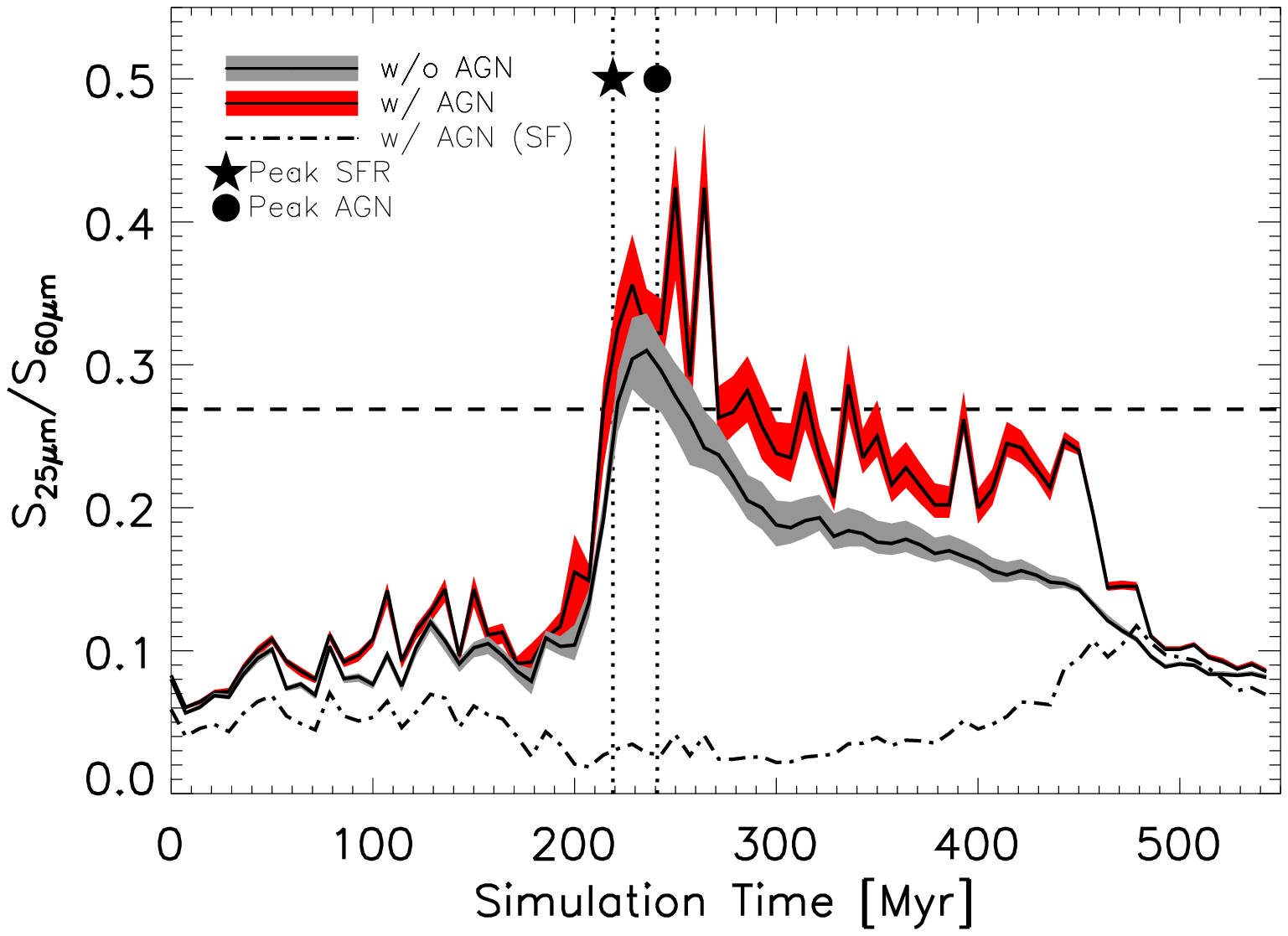,width=3.3in}
\epsfig{figure=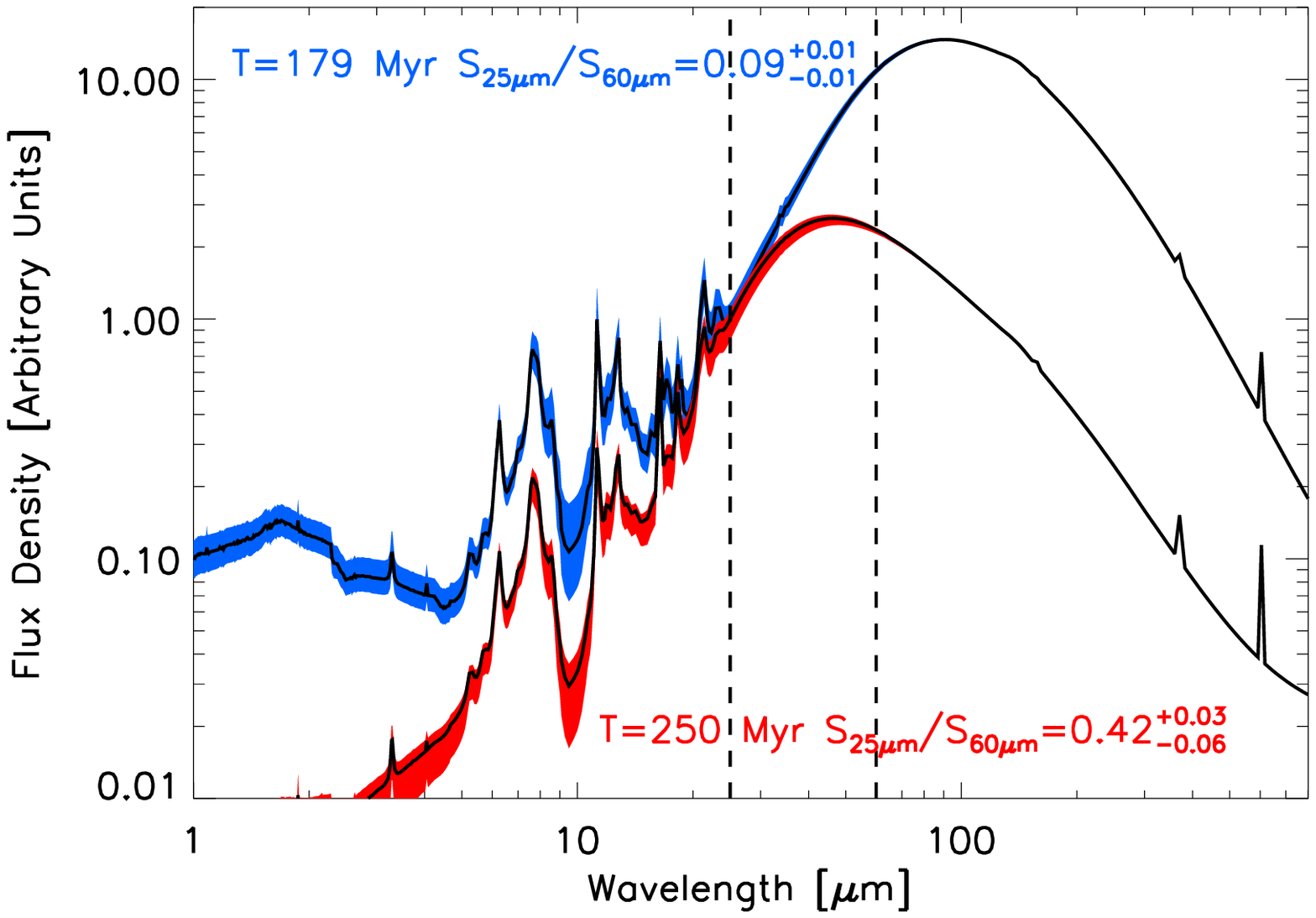,width=3.3in}
\caption{{\sc Left:} Time evolution of the 25--to--60\micron flux density ratio during the final stages of the merger for the e orbital orientation \citep[see \S~\ref{sec:methods} and][]{cox2006} assuming a 20\% covering fraction of PDRs -- cold, dense gas immediately surrounding the young stellar particles -- and a negligible volume filling fraction of molecular gas in the diffuse ISM.  The grey and red shaded areas show the range over six different sightlines when excluding and including the AGN respectively; the black solid line indicates the median value for each.  The dash--dotted black line indicates results including an AGN but assuming a single--phase ISM, which has been shown to maximize the attenuation \citep{witt1996}.  Dotted lines with a star and circle identify the time of peak of star formation and AGN activity respectively, and the dashed black line shows the traditional criterion for warm IR objects $S_{\rm 25\mu m}/S_{\rm 60 \mu m}\gsim 0.27$ \citep[spectral index $\alpha({\rm 25\mu m,60\mu m}) > -1.5$:][]{deGrijp1985,deGrijp1987}.  We find that given a low volume filling fraction of dense molecular gas, our simulations and radiative transfer calculations produce warm (U)LIRGs. {\sc Right:} Examples of the emergent SEDs -- normalized at 25\micronend -- during the cold (blue) and warm (red) evolutionary phases including both AGN and star formation.  The shaded region indicates the range over all sight--lines, and the black solid line shows the mean value.}
\label{fig:warm}
\end{figure*}

Simulation snapshots were post-processed using a significantly
updated version of the polychromatic Monte Carlo radiative transfer
code {\sc Sunrise} \citep[][updates will be described in Jonsson \& Groves, in preparation]{jonsson2006}. 
The adaptive grid resolution was set to 5 kpc allowing for 9 levels of refinement with a lower limit of two 
gas smoothing lengths, which at the peak of the starburst corresponded to $\approx 30$ pc.
We use the R=3.1 Milky Way dust model of \citet{weingartner2001} with a
constant Gas/Dust\footnote{We haven chosen to set a
 constant Gas/Dust ratio, rather than the more physically motivated Metals/Dust, because it is directly constrained by observations of (U)LIRGs  \citep[e.g.,][]{Wilson2008.smaulirgs}, but note that the IR colours for Gas/Dust=50 are consistent with those assuming Metals/Dust=0.4 \citep{dwek1998b} to within 10\%.} ratio of 50 by mass, which is consistent with
observations of local ULIRGs \citep{wilson2008,iono2008.smaulirgs}.
We consider two different treatments of dust in the ISM: (1) the dust density in a given grid cell 
is determined using the multiphase ISM model of \citet{SpringelHernquist2003} assuming a negligible volume filling fraction for dense 
molecular gas -- i.e., only dust in the diffuse phase contributes to the attenuation --  
and (2) a single--phase ISM in which the dust is uniformly distributed in the grid cells -- a configuration that 
has been shown to maximize the attenuation \citep{witt1996}.  
Furthermore, we treat the effects of dust self-absorption properly by iterating the
dust emission and absorption stages of the radiative transfer
calculation until the dust temperature converges \citep{misselt2001}. 
Jonsson \& Groves (in preparation) find that their implementation recovers 
the benchmark 2--D radiative transfer solution of \citet{pascucci2004} to better than $\sim$ few \%
for UV through millimeter wavelengths. 
Finally, we consider the emergent IR SED over 6
different sight lines, but note that at $\lambda>20\mu$m it is fairly
isotropic (see Figure~\ref{fig:warm}).

We assign the stellar particles
SEDs from the library of \citet{groves2008}, which accounts for the
effects of the HII regions and photodissociation regions (PDRs)  --  the transition layer between the HII region and the
surrounding molecular cloud -- that
surround young star clusters. \citet{groves2008} use input spectra computed
using the Starburst99 \citep{leitherer1999} stellar population
synthesis code with a \citet{kroupa2001} initial mass function (IMF). They
calculate the transfer of line and continuum radiation, including the
effects of both gas and dust, through the surrounding HII region
\citep[which evolves as a one-dimensional mass-loss bubble following][]{castor1975} using the {\sc Mappings} III code \citep{groves2004}. The HII regions absorb most of the ionizing radiation and are responsible for
essentially all the hydrogen line emission and the hottest dust
emission. As a subresolution treatment of dense molecular gas, 
 a fraction $f_{\rm pdr}$  -- which is related to the cloud clearing timescale as $f_{\rm pdr} = \exp{(-t/\tau_{\rm clear})}$ \citep[see][]{groves2008} --
 of the emergent luminosity of young stellar clusters are surrounded by PDRs.  The PDRs absorb a large fraction of the
non-ionizing UV photons and potentially contribute significantly to the PAH and
FIR emission; this parameter encodes the effects of obscuration and reprocessed emission owing to cold, dense clouds 
immediately surrounding the young star--forming regions. 

Finally, the SMBH particles are assigned an intrinsic SED from
the \citet{hopkins2007.templateqso} library of unobscured quasar templates
according to their bolometric luminosities, $L_{AGN} = \epsilon \dot{M}_{BH} c^2$,
where $\dot{M}_{BH}$ is the accretion rate and $\epsilon = 0.1$ is the radiative
efficiency corresponding to a standard \citet{shakura1973} thin
disk.  This empirical model SED includes absorption and emission -- including the torus 
contribution -- arising on scales below the resolution of the {\sc Sunrise} grid ($\lsim 30$ pc, see above).
It does not, however, include significant additional obscuration on those scales from, e.g., an 
intrinsically Compton--thick AGN.

\section{Results and Discussion}
\label{sec:results}

\begin{figure}
\epsfig{figure=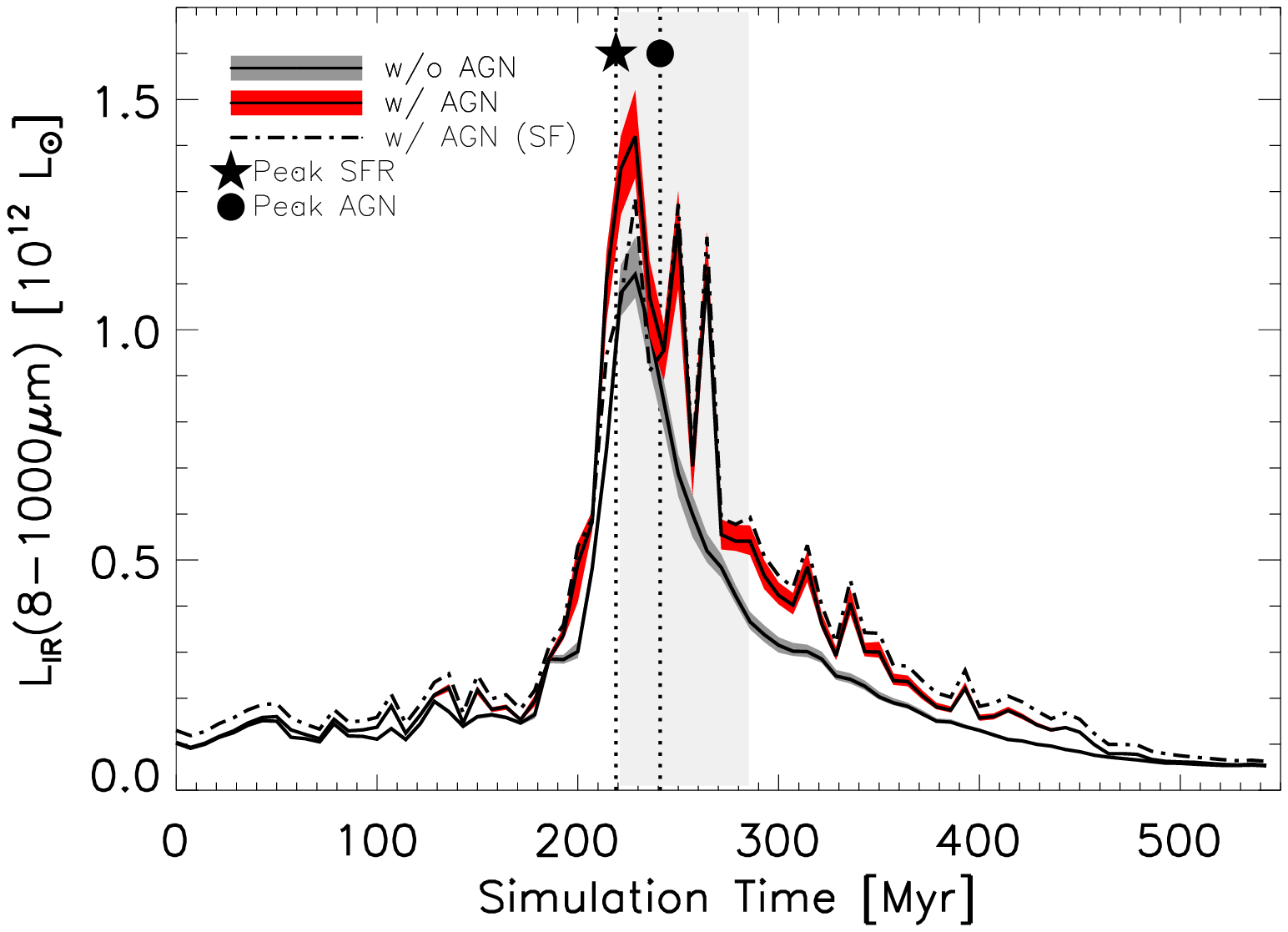,width=3.5in}
\caption{Same labeling as Figure~\ref{fig:warm} for the total infrared luminosity \citep[$8 - 1000{\rm \mu m}$, as defined in ][]{sanders1996}, also for the e orbital orientation.  The shaded region indicates when the object satisfies the warm criterion of \citet{deGrijp1985,deGrijp1987}, and it is considered a ULIRG at any time when $L_{IR} > 10^{12}$ $L_\odot$.  We find that the simulated object is warm during the ULIRG phase.  Furthermore, we find that at times the AGN contributes up to $\approx 30-50\%$ of the peak IR luminosity during this phase.}
\label{fig:lir}
\end{figure}

In Figure~\ref{fig:warm} we present the 25--to--60\micron flux density ratio typically used to distinguish warm (U)LIRGs, including runs both with and without the AGN contribution, and in Figure~\ref{fig:lir} we show the total IR luminosity \citep[as defined in][]{sanders1996} as a function of simulation time.  We choose a fiducial value of $f_{\rm pdr}=0.2$ -- representative of dense molecular gas around young stellar clusters with a $20\%$ covering fraction -- which is roughly consistent with some observations of the nuclear starbursts in ULIRGs \citep[e.g.,][]{downes1998}.  We find that, given this modest covering fraction, while the AGN contributes additional heating, our simulations produce a warm (U)LIRG even when only stellar emission is considered.  Therefore, we find that that stellar emission alone may be sufficient to produce warm IR colours while this system is a ULIRG.  

\begin{figure}
\epsfig{figure=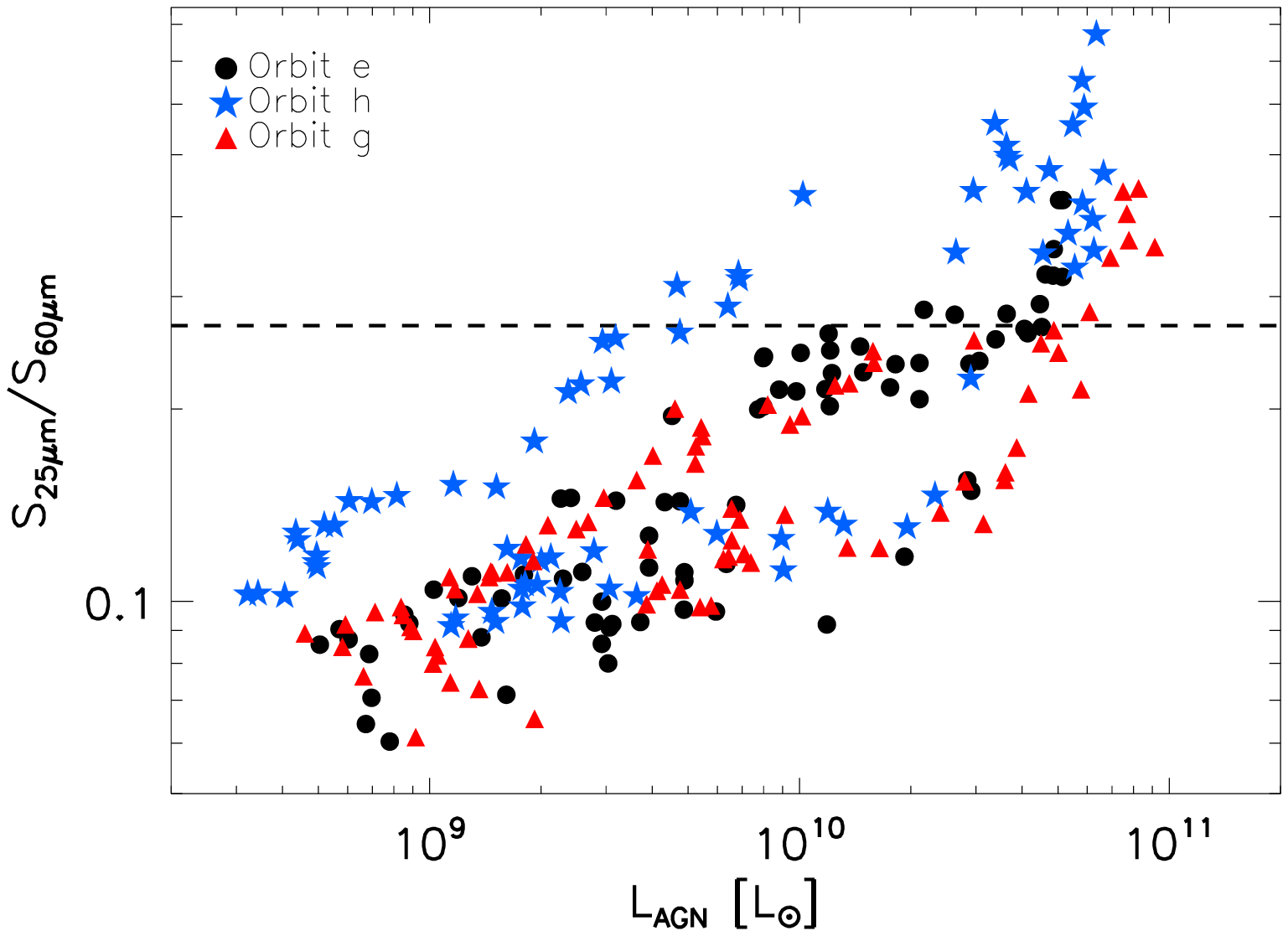,width=3.5in}
\caption{The 25--to--60\micron flux density ratio as a function of AGN luminosity assuming a radiative efficiency of $\epsilon=0.1$ corresponding to standard \citet{shakura1973} thin disk accretion. Included are results from three different orbital orientations \citep[see \S~\ref{sec:methods} and][]{cox2006}: configurations e (black circle), h (blue star), and g (red triangle).  The dashed black line shows the traditional cutoff for warm IR objects $S_{\rm 25\mu m}/S_{\rm 60 \mu m}\gsim 0.27$ \citep[spectral index $\alpha({\rm 25\mu m,60\mu m}) > -1.5$:][]{deGrijp1985,deGrijp1987}.  We find a strong positive correlation between $S_{\rm 25\mu m}/S_{\rm 60 \mu m}$ and $L_{AGN}$ in all cases, but with significant scatter about the mean trend.  The significant scatter in this scaling indicates that while the warm phase is broadly correlated with increased AGN activity, a number of factors are significant including UV emission from young stellar populations and the spatial distribution of dust.}
\label{fig:agn}
\end{figure}

These results are consistent with broadly attributing the warming of the IR SED to AGN activity.  However, they also demonstrate that  warm IR sources do not necessarily contain a significant contribution from an AGN.  Rather, these objects necessarily contain a mix of AGN and star formation that both contribute UV photons that, after being reprocessed by the intervening ISM, lead to a hot dust component at $25 \lsim \lambda \lsim 60$\micron in their SED.  Figure~\ref{fig:agn} shows this explicitly: while there is a strong correlation between $S_{\rm 25\mu m}/S_{\rm 60\mu m}$ and the intrinsic AGN luminosity ($L_{AGN}$) -- a rank correlation coefficient analysis shows that orbits e, h and g have highly significant ($>6\sigma$) strong positive correlations of $\rho = 0.8$, 0.8, and 0.9 respectively -- there is also significant scatter about the mean trend.  This scatter is indicative of contributions of star formation to the warming of the IR SED.  Furthermore, the 25--60\micron colours are produced by the large--scale ISM at all times; even if all of the intrinsic AGN luminosity were reprocessed on small scales and reemitted at 25--60\micron, it would still only contribute 10--15\% of the emergent luminosity at those wavelengths.  

At the same time, the maxima in $S_{\rm 25\mu m}/S_{\rm 60\mu m}$ do not occur at the absolute peak of AGN activity.  In our simulations, the emission at 25 and 60\micron is always dominated by dust in the diffuse ISM.  Therefore, these simulated warm IR colours cannot be attributed directly to the intrinsic AGN emission.  Unfortunately, the complex geometry of the obscuring material and source distribution make it difficult to isolate the origin of this effect, and a number of processes may contribute.  However, it may owe in part to a combination of dust self--absorption and gas depletion.  During the peak of AGN activity, the gas in the nucleus is very dense and may be optically thick to its own emission; this will tend to lower the temperature despite a higher input of UV photons.  At the same time, gas depletion due to the nuclear starburst and outflows driven by AGN feedback will tend to reduce the dust mass in the nucleus after AGN activity peaks.  Since the system will still reprocess almost all of the input UV/optical into the IR, and the IR luminous scales roughly as $L_{IR}\sim M_d T_d^{4+\beta}$ \citep[where $\beta\approx 1.5$ is the dust emissivity; see e.g.,][]{DeBreuk2003,younger2008.egsulirgs}, this will tend to raise the dust temperature at later times.  

We have also performed some preliminary tests in which we assume a single--phase ISM (see Figures~\ref{fig:warm} and \ref{fig:lir}), which has been shown to maximize the attenuation \citep{witt1996}.  In this scenario, the combined effects of a higher dust mass -- which requires a colder effective dust temperature at fixed $L_{IR}$ (see above) -- and self--absorption -- when the dust is optically thick to its own emission, also leading to lower effective dust temperature -- dramatically suppress warm IR colours both with and without an AGN contribution while reaching very similar total IR luminosities; in this maximal attenuation case, even the AGN is insufficient to produce warm colours.  Though a more sophisticated implementation of cold dense gas in the ISM is warranted \citep[e.g.,][]{narayanan2008.winds,li2008}, these preliminary tests suggest that the production of warm IR colours is very sensitive to the effects of attenuation by molecular gas along the line of sight.  Furthermore, while assuming a single--phase ISM may seem an extreme limiting case, observations of local ULIRGs indicate that the covering fractions of optically thick molecular gas in the nuclear starbursts powering local ULIRGs may be close to unity \citep[e.g.,][]{sakamato1999}.  This potentially complicates the interpretation of IR colours as an indicator of AGN activity.

\begin{figure}
\epsfig{figure=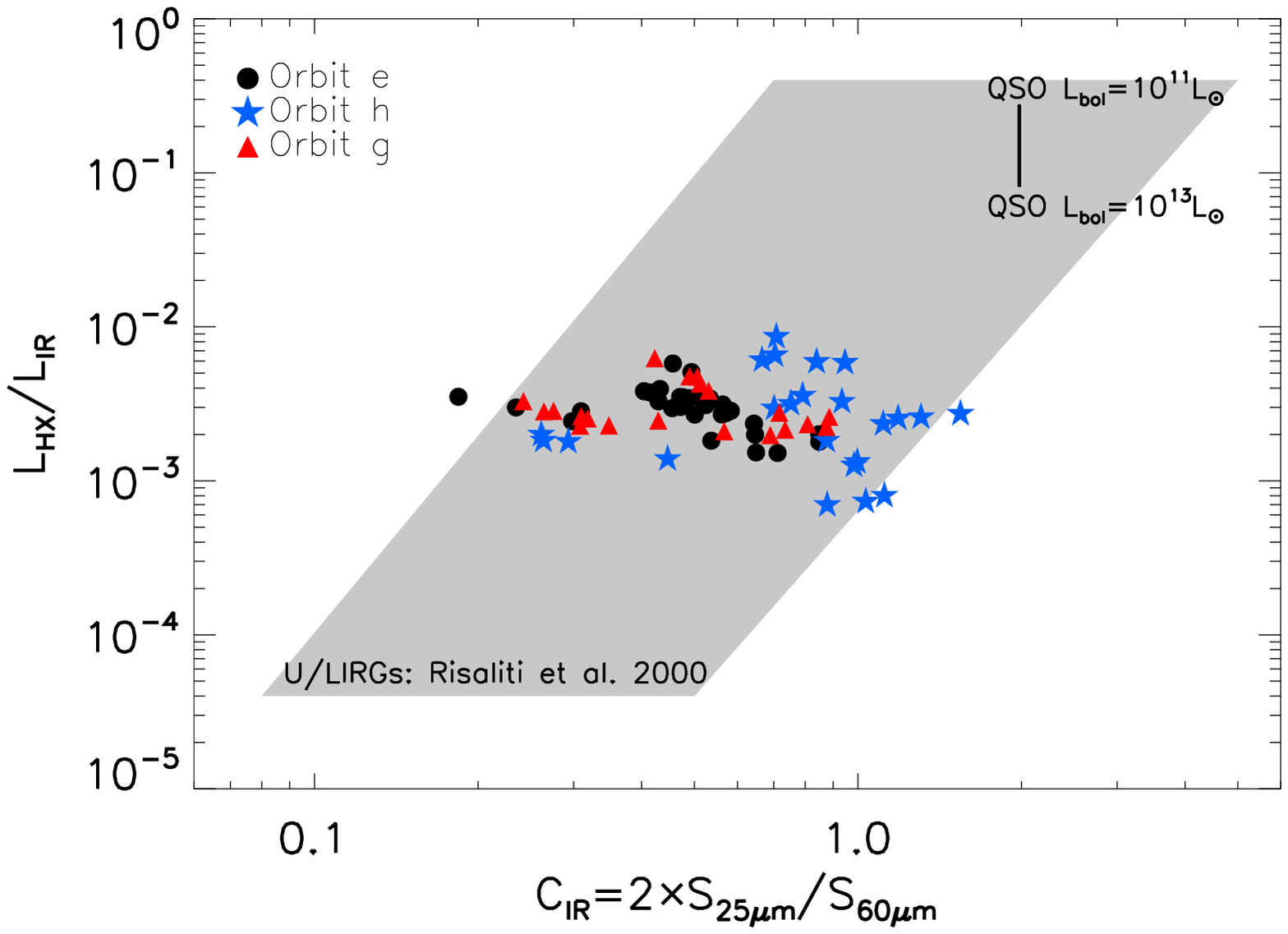,width=3.5in}
\caption{Correlation between $L_{HX}/L_{IR}$ -- where $L_{HX}$ is the hard X--ray \citep[2-10keV; estimated assuming a radiative efficiency of 10\% and the intrinsic AGN SED library of][]{hopkins2007.templateqso} and $L_{IR}$ is the total (8-1000$\mu$m) IR luminosity -- and $C_{IR}=2\times S_{\rm 25\mu m}/S_{\rm 60 \mu m}$ for our simulations compared to the observations of \citet{risaliti2000}.  The grey shaded area shows the range in observed systems from \citet{risaliti2000}.  Shown are results from three orbital configurations \citep[see \S~\ref{sec:methods} and][]{cox2006} -- e (black circle), h (blue star) and g (red triangle) -- and $f_{\rm pdr}=0.2$.  We furthermore restrict ourselves to times when the simulated systems are at $>0.25\times {\rm max(SFR)}$ or $>0.25\times {\rm max}(L_{AGN})$ -- though this does not significantly change the region occupied by our simulations -- to exclude times when the progenitor galaxies are widely separated; our models concern merger--driven starbursts rather than isolated disks.  Finally, we also indicate the location of unobscured quasar templates from \citet{hopkins2007.templateqso} for $L_{AGN}=10^{11-13}$ $L_\odot$.  We find that our merger--driven evolutionary models of warm (U)LIRGs are consistent with the observed scatter in 25--to--60\micron flux density ratio at fixed $L_{HX}/L_{IR}$.}
\label{fig:hx}
\end{figure}

To test the predictions of our models observationally, we compare our simulated systems -- including two additional orbital configurations -- with hard X--ray (HX: 2--10 keV) observations of analogous objects \citep[see Figure~\ref{fig:hx}:][]{risaliti2000}.  We find that our simulations -- assuming a radiative efficiency of 10\% and the intrinsic SED library of \citet{hopkins2007.templateqso} for the AGN -- are consistent with the observed scatter in $C_{IR}$ within a narrow range in $L_{HX}/L_{IR}$.  The relatively small dynamical range of $L_{HX}/L_{IR}$ seen in our simulations is not surprising; we have only sampled a single mass encounter over a small number of orbital configurations and parameter choices.  Indeed, a thorough study of the effects of varying the Gas/Dust ratio, dense molecular gas along the line--of--sight, galaxy/halo mass, and others are all topics we plan to study in detail in future work.  Furthermore, there could be additional obscuration near the AGN on scales below our spatial and mass resolution or owing to dense gas along the line of sight would yield significantly higher column densities\footnote{In our simulations, the optical depth of the HX to photoelectric absorption \citep{Morrison1983} is $\tau < 0.05$ at all times of interest.} similar to those which have been observed locally \citep[e.g.,][]{maiolino1998,bassani1999} which would tend to lower both $L_{HX}/L_{IR}$ and $S_{\rm 25\mu m}/S_{\rm 60 \mu m}$.  Conversely, lower column densities along particular sight lines could tend towards higher $L_{HX}/L_{IR}$ and $S_{\rm 25\mu m}/S_{\rm 60 \mu m}$.  Finally, gas--rich isolated star--forming disks with radiatively inefficient or low--level AGN \citep[e.g.,][]{hopkinshernquist2006,hopkinshernquist2008a,hopkinshernquist2008b} could have low $L_{HX}/L_{IR}$ and a range of $S_{\rm 25\mu m}/S_{\rm 60 \mu m}$ colours.  However, our models of a merger--driven evolutionary scenario for the production of warm (U)LIRGs are consistent with the constraints imposed by the observations of \citet{risaliti2000} and others.  However, at the same time our simulations demonstrate that  the correlation between $C_{IR}$ and $L_{HX}/L_{IR}$ should not be taken to show a direct link between warm colours and AGN activity; rather, their origin is inherently ambiguous, produced by star formation, AGN activity, or both in combination.

\section{Conclusion}
\label{sec:conclude}

We present the results of radiative transfer calculations, including a self--consistent treatment of dust heating, on simulations of gas--rich major mergers.  Our models produce warm IR SEDs in systems analogous to local (U)LIRGs, and we find that the the warm phase is broadly correlated with AGN activity.  Furthermore, the mid--IR emission arises almost entirely from the diffuse ISM rather than the intrinsic AGN SED.  But, we show that star formation alone may be sufficient to produce warm IR colours, which suggests that they may be a more ambiguous probe of AGN activity.   However, these results are sensitive to the multiphase structure of the ISM; when we consider a single--phase ISM, which maximizes the obscuration \citep{witt1996}, even a significant AGN contribution will not produce warm IR colours.  Finally, we find that the HX and IR properties -- and specifically the observed scatter in the 25--to--60\micron flux density ratio at fixed $L_{HX}/L_{IR}$ -- of local (U)LIRGs are consistent with our models of a merger--driven evolutionary scenario in which both star formation and AGN activity probably contribute significantly to the production of warm (U)LIRGs, and that these objects are intermediate between the star formation dominated cold phase and the birth of a bright, unobscured quasar.

\section*{Acknowledgements}

We thank Giovanni G. Fazio, Stephanie J. Bush, Dusan Keres, and Phillip F. Hopkins for helpful discussions.  The simulations and radiative transfer calculations presented here were performed on the Harvard FAS High Performance Computing Cluster in Boston, MA.  PJ was supported by programs HST-AR-10678 and 10958, provided by NASA through a grant from the Space Telescope Science Institute, which is
operated by the Association of Universities for Research in Astronomy, Incorporated, under NASA contract NAS5-26555, and by Spitzer Theory
Grant 30183 from the Jet Propulsion Laboratory.  CCH acknowledges support from a National
Science Foundation Graduate Research Fellowship. This work was supported in part by a grant from the W. M. Keck Foundation.

\bibliographystyle{apj}
\bibliography{../../smg}

\end{document}